\documentclass{PoS}

\usepackage{lineno} 

\usepackage{graphics,graphicx} 

\usepackage{wrapfig} 

\usepackage[caption=false]{subfig} 

\title{
TeV Gamma-Ray Emission Observed from Geminga with HAWC
}

\ShortTitle{
TeV Gamma-Ray Emission Observed from Geminga with HAWC
}

\author{ B.~M.~Baughman$^a$ and \speaker{J.~Wood}$^a$ 
        for the HAWC Collaboration$^b$ \\
        \llap{$^a$}Department of Physics, University of Maryland, 
        College Park, MD, USA \\
        \llap{$^b$}For a complete author list, see \href{http://www.hawc-observatory.org/collaboration/icrc2015.php}{www.hawc-observatory.org/collaboration/icrc2015.php}.        
        Email: \email{jwood@umdgrb.umd.edu}, \email{bbaugh@umdgrb.umd.edu}}

\abstract{
Geminga is a radio-quiet pulsar ~250 parsecs from Earth that was first 
discovered as a GeV gamma-ray source and then identified as a pulsar.
Milagro observed an extended TeV source spatially consistent with Geminga.
HAWC observes a similarly extended source.
Observations of Geminga's flux and extension will be presented.
}

\FullConference{The 34th International Cosmic Ray Conference,\\
		30 July- 6 August, 2015\\
		The Hague, The Netherlands}

\begin{document}

\section{Introduction}
Geminga is an important high energy source.
It is one of the closest known middle aged pulsars~\cite{citeulike:2148690} at 
approximately 250 pc~\cite{citeulike:1436021}.
It was discovered in $\gamma$-rays with the SAS-2 
experiment~\cite{citeulike:13663335} before being observed in other wavelengths.
Pulsations were first observed in X-rays~\cite{citeulike:13663336} with ROSAT and
later EGRET~\cite{citeulike:13663337} but have not been observed in the 
TeV range~\cite{citeulike:13663341,citeulike:5330366,citeulike:13663339}.
X-ray Multimirror Mission-Newton observations show a pulsar wind around Geminga 
with an extent of a few arc minutes~\cite{citeulike:13663300}.
Observations of TeV emission associated with Geminga would bolster the 
interpretation that Geminga is a nearby cosmic-ray 
accelerator~\cite{citeulike:13665961,citeulike:3903568,citeulike:10003405} 
which could possibly explain the observed positron 
excess~\cite{citeulike:3486162,citeulike:13148665,citeulike:13669840}.

Milagro reported Geminga as a possible TeV source in 
2007~\cite{citeulike:12338581} 
with a significance of $5.1 \sigma$ before trials but 
below discovery threshold after trials.
In 2009 Milagro made a ``definitive detection'' of a region of extended 
${(2.6^{+0.7}_{-0.9})}^\circ$ emission spatially consistent with 
Geminga~\cite{citeulike:12338580} with a significance of $6.3 \sigma$.
This would imply a region of emission approximately 6 to 12 pc in 
extent.
The Tibet Air shower array reported an excess of $2.2 \sigma$ at the location
of the pulsar but did not report extended emission~\cite{citeulike:8023909}.

Imaging Atmospheric Cherenkov Telescopes (IACTs) have
observed Geminga without 
significant 
detection~\cite{citeulike:13663341,citeulike:13663330,citeulike:5330366} for 
over two decades.
An extended, hard TeV source is extremely difficult to observe with the IACT 
technique.
Milagro's observation implies an extended, hard spectrum source.

HAWC was inaugurated on March 20th, 2015 with over 250 Water Cherenkov 
Detectors (WCDs) and a predicted point source sensitivity of 10 times that of 
its predecessor Milagro.
The water Cherenkov technique allows HAWC to have a nearly 100\% duty cycle 
and large field of view, making the HAWC observatory an ideal instrument for 
the study of transient phenomena and able to see large scale diffuse features 
in the TeV sky.
In this proceeding we describe the analysis of gamma rays TeV from the Geminga 
remnant using the 250-tank configuration of the HAWC array, hereafter referred
to as HAWC-250.
\begin{figure}
  \centering
  \subfloat[Milagro]{\includegraphics[width=0.49\textwidth ]
  {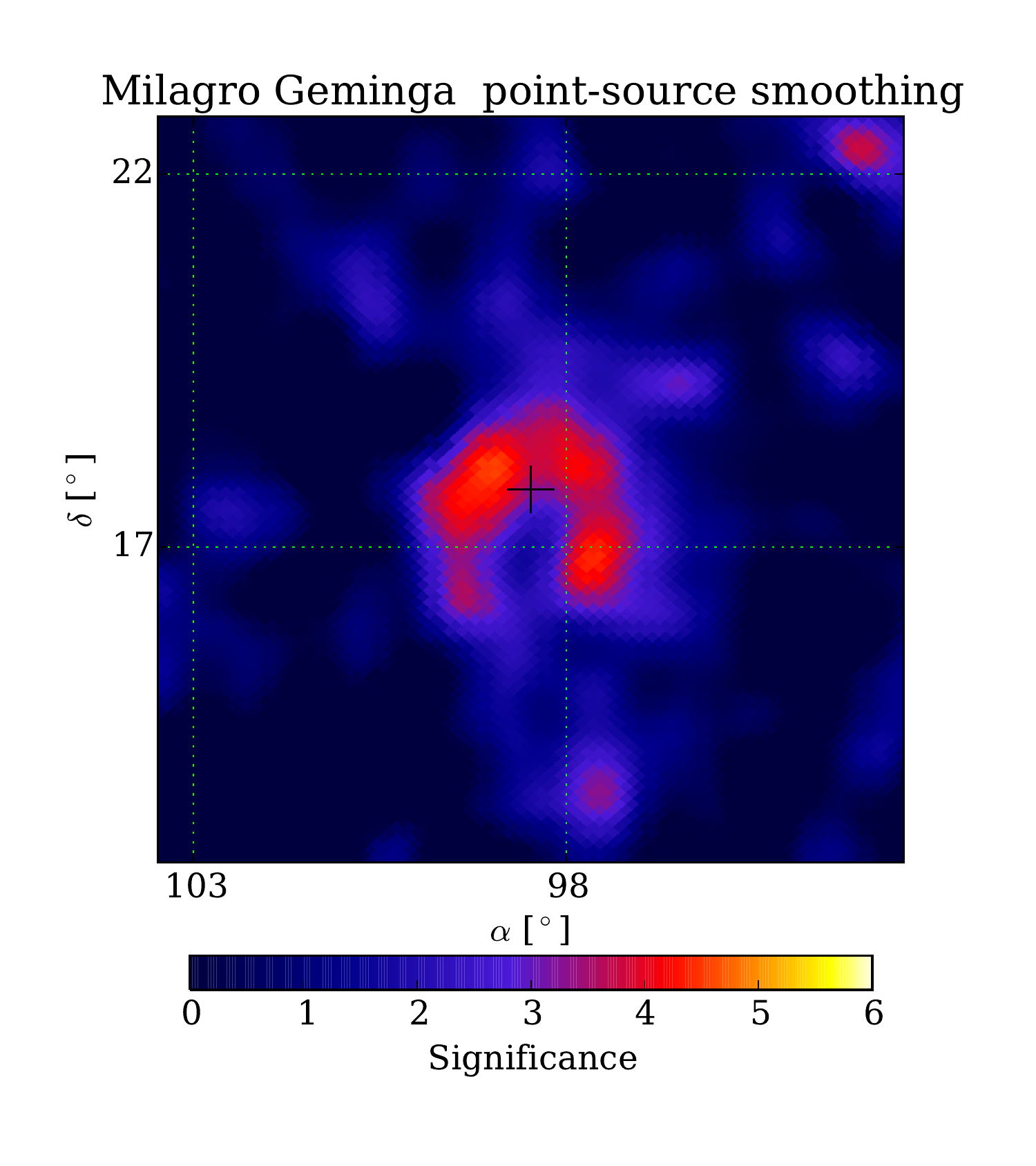}}
  \subfloat[HAWC]{\includegraphics[width=0.49\textwidth]
  {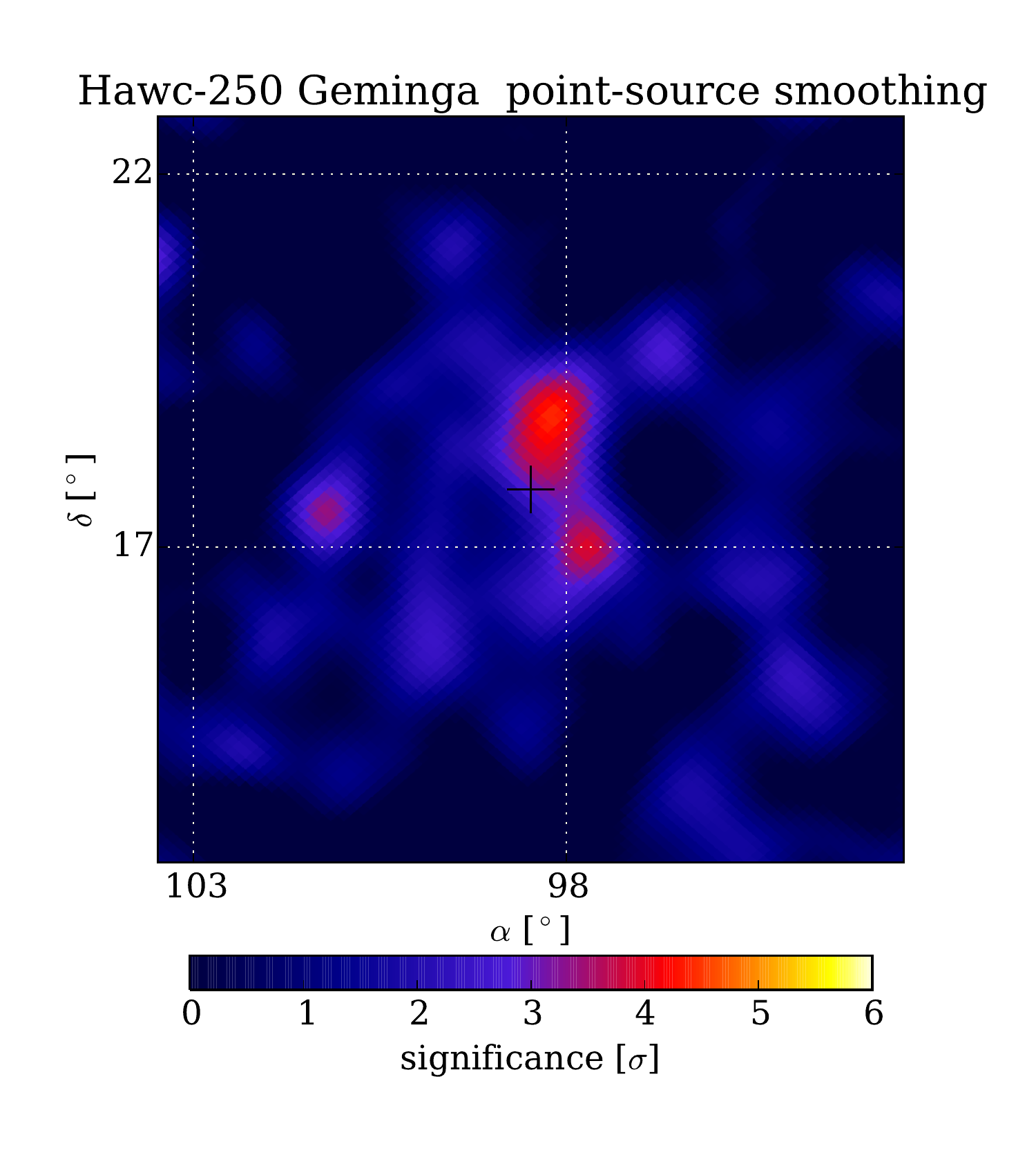}}
  \caption{\footnotesize The significance for a 
  10$^\circ$x10$^\circ$ region around Fermi source J0634.0+1745,
  the Geminga pulsar.
  HAWC's energy threshold is significantly lower than that of Milagro's for
  these data.
  }
  \label{fig:pnt-obs}  
\end{figure}

\begin{figure}
  \centering
  \subfloat[Milagro]{\includegraphics[width=0.49\textwidth ]
  {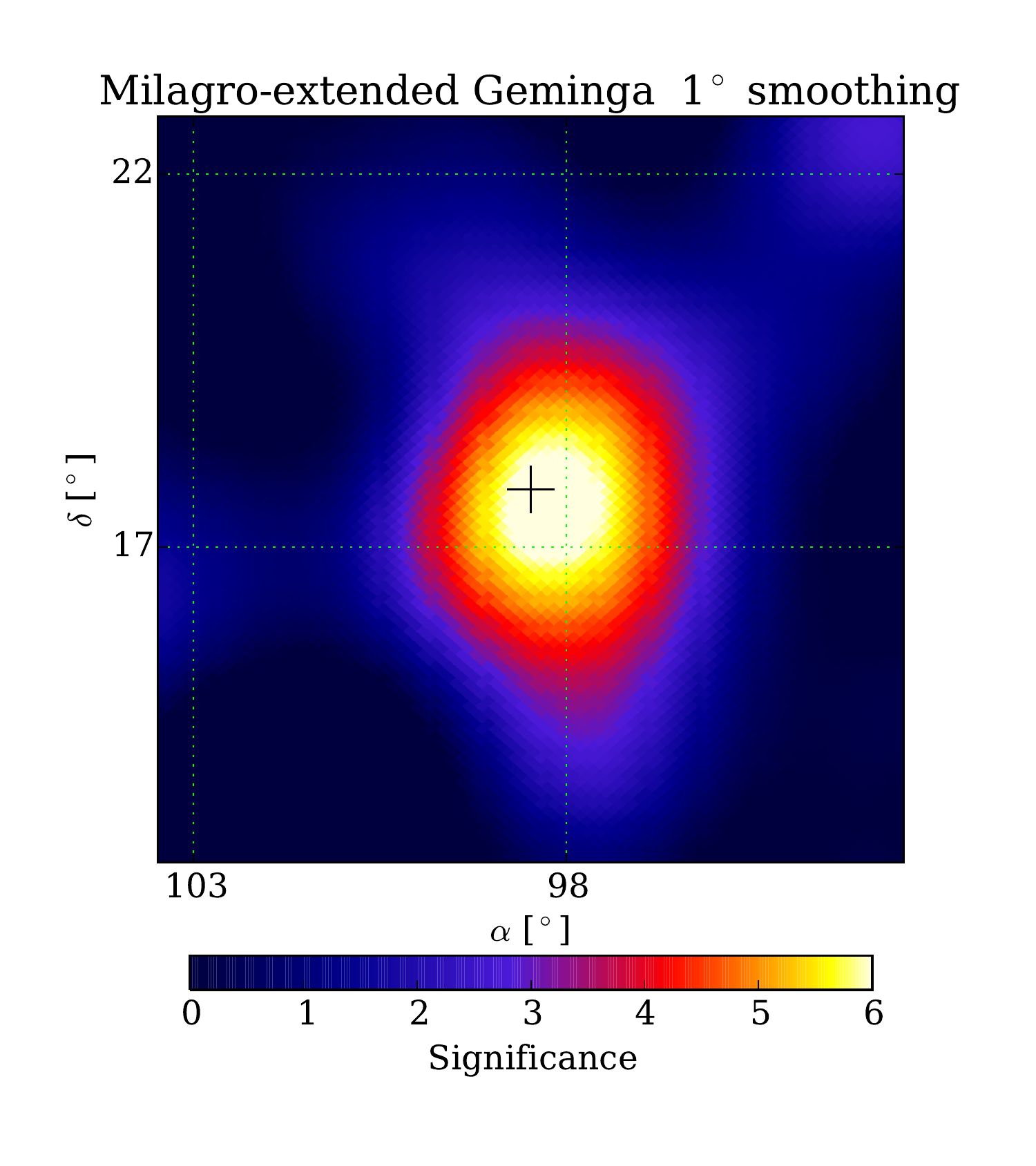}}
  \subfloat[HAWC]{\includegraphics[width=0.49\textwidth]
  {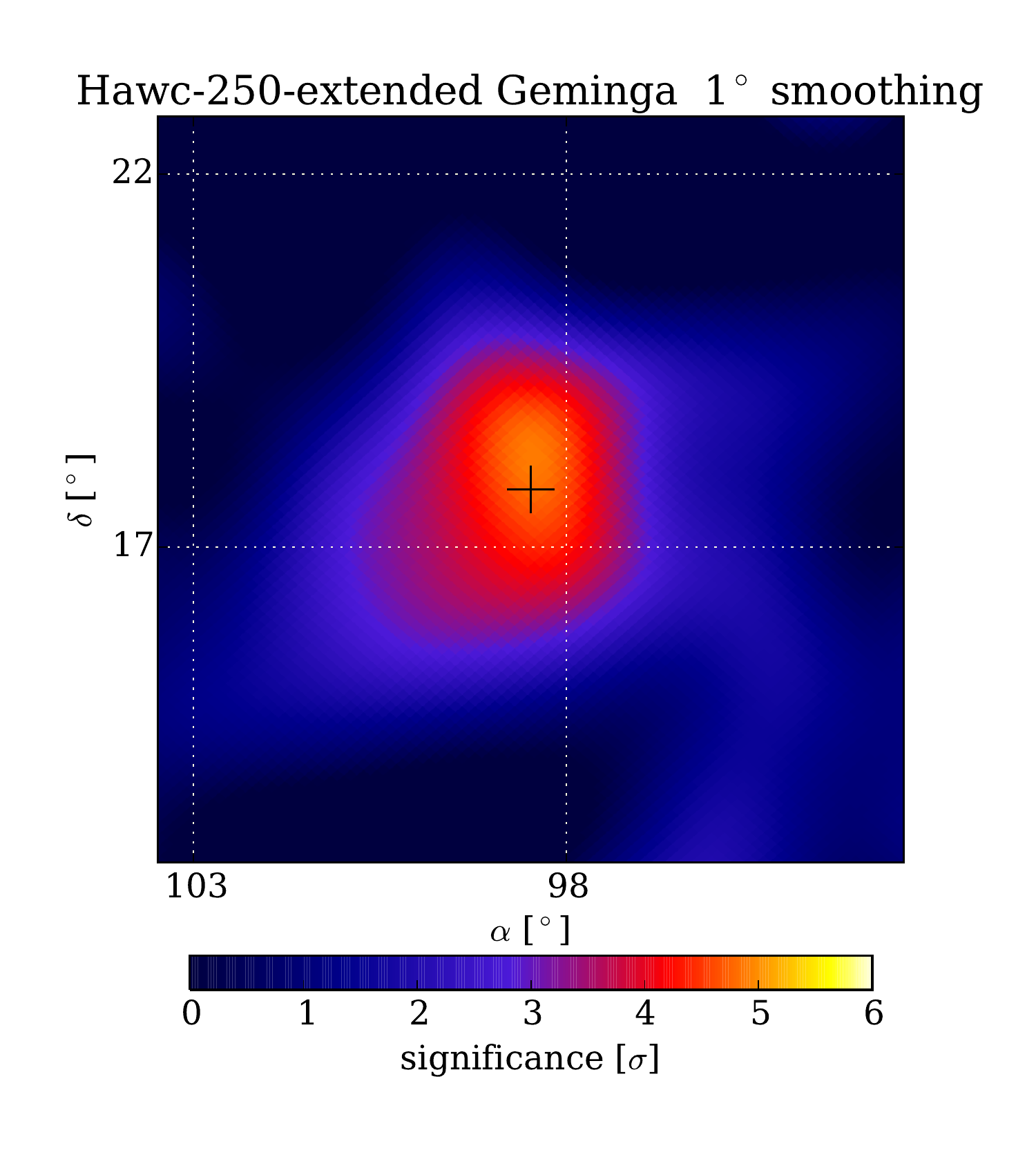}}
  \caption{\footnotesize The significance after
  smoothed by an additional 1$^\circ$ Gaussian of a 
  10$^\circ$x10$^\circ$ region around Fermi source J0634.0+1745,
  the Geminga pulsar.
  HAWC's energy threshold is significantly lower than that of Milagro's for
  these data.
  }
  \label{fig:ext-obs}  
\end{figure}

\section{Results}
In this paper we present HAWC's preliminary results showing an extended 
region of emission consistent with that observed with Milagro.
HAWC-250, where the number of operating water Cherenkov detectors ranged from 
247 to 293, data were used for this analysis.
Data taken from November 26th 2014 to May 6th 2015 totaling to a live-time of
149 days\footnote{See ~\cite{Salesa:2015} for more details on this data set.}.
These data result in a $\sim 38 \sigma$ detection of the 
Crab~\cite{Salesa:2015}.
Milagro, for comparison, reported $\sim 17 \sigma$ on the Crab in the same
data set used for their Geminga analysis~\cite{citeulike:12338580}.

HAWC gains much of its sensitivity improvement over Milagro from its improved
point spread function~\cite{Salesa:2015}.
Extended objects do not benefit as much from this improvement as their flux is 
spread over a larger solid angle.
HAWC's low energy sensitivity is also greatly improved.
At high energies, HAWC and Milagro have similar effective areas. 
Thus, hard spectrum, extended sources do not benefit from the improved low 
energy sensitivity of HAWC. 

The region around Geminga can be seen for both HAWC and Milagro optimized for 
point source detections\footnote{
Descriptions of HAWC's and Milagro's point source analyses can be found 
in~\cite{Salesa:2015} and ~\cite{citeulike:12338580} respectively.
}
in Fig.~\ref{fig:pnt-obs}, while Fig.~\ref{fig:ext-obs}
shows the same data with an additional $1^{\circ}$ smearing to highlight 
extended sources.

HAWC has exceeded the integrated point-source sensitivity of its predecessor,
Milagro, by over a factor of two in just 149 days of live-time and confirms
Milagro's observation of an extended TeV source spatially 
coincident with the Geminga pulsar.
It  will continue to update the community on this extremely interesting source 
over its lifetime.

\acknowledgments{}
We acknowledge the support from: the US National Science Foundation (NSF);
the US Department of Energy Office of High-Energy Physics;
the Laboratory Directed Research and Development (LDRD) program of
Los Alamos National Laboratory; 
Consejo Nacional de Ciencia y Tecnolog\'{\i}a (CONACyT),
Mexico (grants 260378, 55155, 105666, 122331, 132197, 167281, 167733);
Red de F\'{\i}sica de Altas Energ\'{\i}as, Mexico;
DGAPA-UNAM (grants IG100414-3, IN108713,  IN121309, IN115409, IN111315);
VIEP-BUAP (grant 161-EXC-2011);
the University of Wisconsin Alumni Research Foundation;
the Institute of Geophysics, Planetary Physics, and 
Signatures at Los Alamos National Laboratory;
the Luc Binette Foundation UNAM Postdoctoral Fellowship program.

\bibliography{icrc2015-geminga}

\end{document}